\documentclass[english]{article}
\usepackage{rotfloat}
\usepackage{amsmath}
\usepackage{cite}
\usepackage{longtable}
\usepackage{tabularx}
\usepackage{float}
\usepackage{color, colortbl}
\usepackage[letterpaper, margin=1in]{geometry}
\usepackage[font=small,labelfont=bf]{caption}
\usepackage{array}

\usepackage[utf8]{inputenc}
\pdfoutput=1
\usepackage{hyperref}
\hypersetup{
  pdfinfo={
    Title={Your Title Here},
    Author={Your Name Here},
    Subject={If you want to put something here, do so},
    Keywords={Add some keywords if you feel so inclined}
  }}
\usepackage{pdfpages}

\makeatletter
\title{On the Vibrational Model of $^{92}$Pd and Comparison with $^{48}$Cr}

\author{Shadow JQ Robinson$^{1}$, Castaly Fan$^{2}$, \\
Matthew Harper$^{2}$, Larry Zamick$^{2}$\\
\vspace{1.4em}
\\
$^{1}$Department of Chemistry and Physics, University of Tennessee at Martin, \\
Martin, Tennessee 38238, USA\\
$^{2}$Department of Physics and Astronomy, Rutgers University,\\
Piscataway, New Jersey 08854, USA}

\begin{document}
\maketitle

\begin{abstract}
    It was found experimentally that the yrast $I=$ 2$^+$, 4$^+$ and 6$^+$ states of $^{92}$Pd are almost equally spaced suggesting a vibrational band. We consider to what extent this should be considered as an evidence of maximum $J$-pairing for this nucleus.
\end{abstract}

\section{Introduction}

    In a Nature article by Cedewall et al.\cite{1} found near equally spaced yrast $I = 2^{+}, 4^{+}$ and $6^{+}$ states of $^{92}$Pd i.e. a vibrational spectrum. The B(E2)'s were not measured and at present are not known. There is an accompanying article with theoretical support \cite{2} where in part B(E2)'s were calculated and the importance of odd spin $T = 0$ pairing was proposed to explain the results. In a recent work \cite{3} we lent theoretical supports for this picture by noting that in the vibrational model the static quadrupole moment of the $2^{+}$ should be very small if not zero. Large shell model calculations using the program Antoine \cite{4} supported this. The interactions used were JUN45\cite{5} and jj4b\cite{6}. However, the calculations of B(E2)'s were not in agreement with the predictions of the simple harmonic vibrator model.

    There are some similarities in the analyses made in ref \cite{1} in the $g_{9/2}$ shell and much earlier works in the
    $f_{7/2}$ shell by McCullen, Bayman and Zamick \cite{7}\cite{8} and Ginocchio and French\cite{9}.

    There it was noted that both the $J = 0^{+}$ $T = 1$ and $J = J_{max} = 7$ two-body matrix elements were low
    lying. The input matrix elements of MBZE\cite{10} in MeV from $J = 0$ to $J = 7$ are respectively:

    \begin{center}
        0.0000, 0.6111, 1.5863, 1.4904, 2.8153, 1.5101, 3.2420, 0.6260
    \end{center}

    The wave function coefficients of the lowest $I = 0^{+}$ state with this interaction are:
    
    \begin{center}
        $C(0,0) = 0.7878,\quad C(2, 2) = 0.5615,\quad C(4, 4) = 0.2208,\quad C(6, 6) = 0.1234$
    \end{center}
    
    Here $C(J_{p}, J_{n})$ is the probability amplitude that the protons couple to $J_{p}$ and the neutrons to $J_{n}$. This
    wave function is quite different from one for $J = 0$ $T = 1$ pairing interaction. For such an interaction we obtain $C(2,2) = 0.2152$. This is much smaller than what one obtains with the more realistic interaction. An important point in \cite{7}\cite{8}\cite{9} is that although in the $f_{7/2}$ shell seniority is a good quantum number for identical
    particles, it is badly broken for a system of both protons and neutrons. The rare exception is the $J = 0$ $T = 1$
    pairing interaction.

    For two protons and two neutrons in the $g_{9/2}$ shell, a remarkably similar wave function structure was obtained by the Swedish group. The two-body matrix elements used by Qi et al.\cite{11} from $J =0$ to $J = 9$ are:
    
    \begin{center}
        0.0000, 1.2200, 1.4580, 1.5920, 2.2830, 1.8820, 2.5490, 1.9300, 2.6880, 0.6260
    \end{center}

    The resulting coefficients are:

    \begin{center}
        $C(0,0) = 0.76,\quad C(2, 2) = 0.57,\quad C(4, 4) = 0.24,\quad C(6, 6) = 0.13,\quad  C(8,8) =0.14$
    \end{center}
    
    This is in contrast with a $J =0$ $T =1$ pairing interaction for which $C(2,2) = 0.1741$.

    However the Swedish group goes beyond what was done in the past. They attempt to give a physical interpretation of the experimentally observed vibrational spectrum in terms of angular differences of neutron proton pairs coupled to maximum angular momentum e.f $J=9$ in the $g_{9/2}$ shell. There has however been a recent reaction to the over interpretation of the vibrational spectrum in terms of a single $J$ shell model i.e. $g_{9/2}$. Fu Zhao and Arima \cite{12} point out that in the work of Robinson et al. \cite{3} the percentage of pure $g_{9/2}$ configuration in the $J=0^{+}$ ground state and $J=2^{+}$ first excited state is less than $33\%$.

    In this work we will speak of interactions rather than correlations although the two are obviously connected as seen e.g. in recent work of K, Neergaard \cite{13}. Instead of $J=0$ pairing we speak of the $J = 0$ $T = 1$ interaction. Also the $J_{max}$ interaction and the $T = 1$ interaction where all $T = 0$ two-body matrix elements are sent to zero.

    A recent review article contains many references to $J$-pairing in nuclei \cite{14}. It will not be practical to mention them all but we here give a smaller selected list of relevant works, some mentioned in the review and some not \cite{15, 16, 17, 18, 19, 20, 21, 22}

    In the works of A. Goodman \cite{23, 24} it is pointed out that for a given band the $J = 0$ $T = 1$ pairing correlations dominates for low spins, but as one goes up in spin there is a gradual transition so that for the highest spin states $T = 0$ pairing dominates. This can be seen in a simple model of Harper and Zamick \cite{25} for a system of two protons and two neutrons where, for sufficiently high spin, the $J = 0$ $T = 1$ two body matrix element never enters into the calculation. One cannot reach a high value of J if one pair of particles is coupled to zero.

    In this work, we explore whether the effects of $T = 0$ pairing can really be seen in the $^{92}$Pd results \cite{1}. We invoke a modification of the effective interaction as previously used by Robinson and Zamick \cite{26} in the $f$-$p$ shell. There, calculated results were compared when all the two-body matrix elements were present with calculated results where all the $T = 0$ two-body matrix elements are set equal to zero. Such a procedure has been also performed in ref \cite{1} for the $g_{9/2}$ Shell. Undoubtedly, there will be differences with the full interaction and the one where $T = 0$ matrix elements are. set to zero, but will we be able to say anything definitive about $T = 0$ pairing?

    It should be noted that turning the $T = 0$ interaction on and off was previously applied to nuclear binding energies in Satula et al.\cite{27}. They concluded from this study that the $T = 0$ interaction was responsible for the Wigner energy. However, in a similar study Bentley et al. \cite{28} claimed that the $T = 0$ interaction was mainly responsible for the symmetry energy.

    Using this model, Robinson and Zamick \cite{29, 30} found a partial dynamical symmetry when setting $T = 0$ two-body matrix elements to zero in a single 7 shell calculation. For example in $^{44}$Ti $T = 0$ states with angular momenta $I$,where $I$ could not occur in $^{44}$Ca (i.e. could not exist for $T = 2$ because of the Pauli Principle) were degenerate.

\section{Results}
    We present results in Table \ref{tab1} of calculations with the Antoine program \cite{4} using the JUN45 interaction \cite{5}. In the second column we present the calculated energies of the even $I$ states up to $I = 14^{+}$. This is followed by static quadrupole moments and B(E2)'s. We then present results in the third column for what we call T0JUN45. Here we have the same $T = 1$ two-body matrix elements as JUN45 but we set the $T = 0$ two-body matrix elements to zero. This turning  on and off  was previously done for $^{44}$Ti in 2001 by Robinson and Zamick \cite{29}. In the last columns we show the experimental results of Cederrwall et al.\cite{1}.

    We note that the near ``equal spacing'' that was found experimentally up to $J = 6^{+}$, extends much further in the shell model calculation. For example, at $J = 14^{+}$ the calculated energy 5.336 MeV should be compared with 7E(2) = 5.881 MeV.
    
\begin{figure}[H]
    \centering
    \includegraphics[width=9cm]{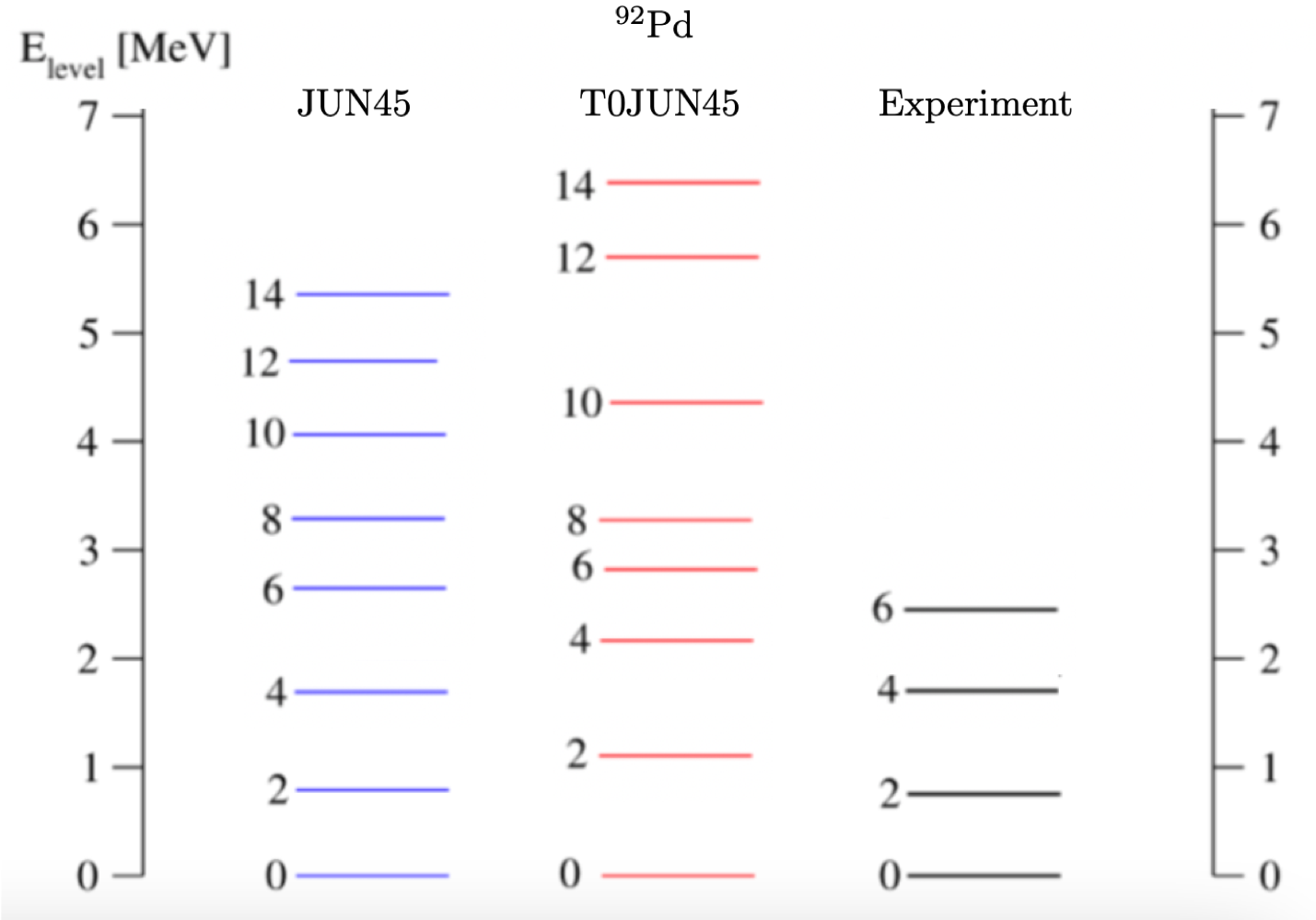}
    \caption{Full $fp$ calculations of even $J$, $T = 0$ states in $^{92}$Pd and comparison with experiment.}
    \label{fig0}
\end{figure}

   In the second column of numbers in Table \ref{tab1}  we have T0JUN45 in which the $T=0$ matrix elements of JUN45 have been set to zero. Note that this has already been addressed by Cederwal et al. They noted that the energy gaps with the full JUN45 interaction were smaller than those with T0JUN45. This is also shown here in Fig \ref{fig0}. This is qualitatively opposite as to what was obtained in reference \cite{29} and as we will soon see for $^{48}$Cr.

    The singular result within the spectra that could be viewed as substantially different between the interactions is found in the $J = 8$ state. This state is found very close to the $J = 6$ state in the calculations where the $T = 0$ matrix elements are removed, but is nearly equally spaced in the calculation utilizing the full interaction. No experimental determination of the location of this crucial state has yet been made.

    We quickly look at the static quadrupole moments.The static quadrupole moment of a state with total angular momentum $J$ is given by
    \begin{equation}
        Q(J) =  \langle \Psi^{J}_{J} | O^{2}_{0} | \Psi^{J}_{J} \rangle
    \end{equation}
    , where
    \begin{equation}
    \begin{split}
        O_{0}^{2} &= \sqrt{\frac{16\pi}{5}}(M_{0p}^{2} + M_{0n}^{2})\\
        M_{0p}^{2} &= e_{\text{eff}}\sum_{\text{protons}}r^{2}(i) Y_{0}^{2}(i)\quad\text{, etc.}
    \end{split}
    \end{equation}
    The standard effective charges $e_{p} =1.5$ and $e_{n}=0.5$ were used for $^{92}$Pd. The quadrupole moments of the $2^{+}$ and $4^{+}$ states are indeed small, especially when compared with quadrupole moments in other nuclei as shown in ref.\cite{26}. It should be noted that in the rotational model a positive laboratory frame quadrupole moment implies that the nucleus is oblate-negative means prolate. The fact that the calculated  quadrupole moments are small  lends support to a vibrational picture for $^{92}$Pd. Note that $Q(2)$ for T0JUN45 is of opposite sign to that of JUN45, $-3.861$ vs $+3.514$. Since the B(E2) involves a square of the matrix element of the quadrupole operator one loses the sign.

    Indeed we now look at the B(E2)'s. For $2 \rightarrow 0$ and $4 \rightarrow 2$ the B(E2)'s with the full interaction are substantially larger than with T0JUN45. The enhancement is due to the $T = 0$ matrix elements in the two-body interaction. But can we use this as evidence? The problem is that there is a large uncertainty in what the effective charges one should use in shell model calculations. One could obtain the enhancement by choosing different effective charges.

    But then if we look at $6 \rightarrow 4$ and $8 \rightarrow 6$ we see big differences. The B(E2)'s for the full cases are 364.14 and 315.074 e$^2$fm$^4$ while with T0JUN45 they are much smaller: 5.017 and 2.741 e$^2$fm$^4$. Perhaps then strong B(E2)'s for $8 \rightarrow 6$ and $6 \rightarrow 4$ are good evidence of the importance of $T = 0$ matrix elements and of partial $J_{max}$ pairing.

\begin{figure}[H]
    \centering
    \includegraphics[width=9cm]{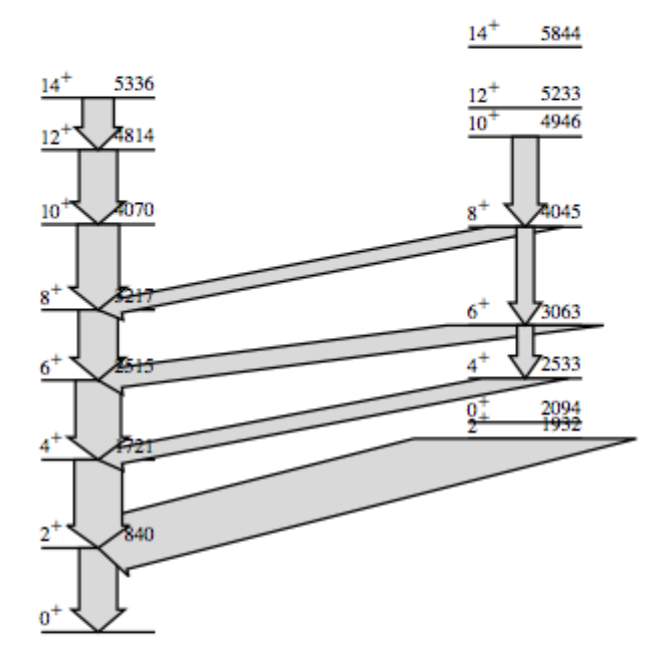}
    \caption{Yrast and Yrare levels $J = 0 - 14$ for $^{92}$Pd using the JUN45 effective interaction. B(E2) values over 100 e$^2$fm$^4$ are shown with width corresponding to strength.}
    \label{fig1}
\end{figure}

\begin{figure}[H]
    \centering
    \includegraphics[width=9cm]{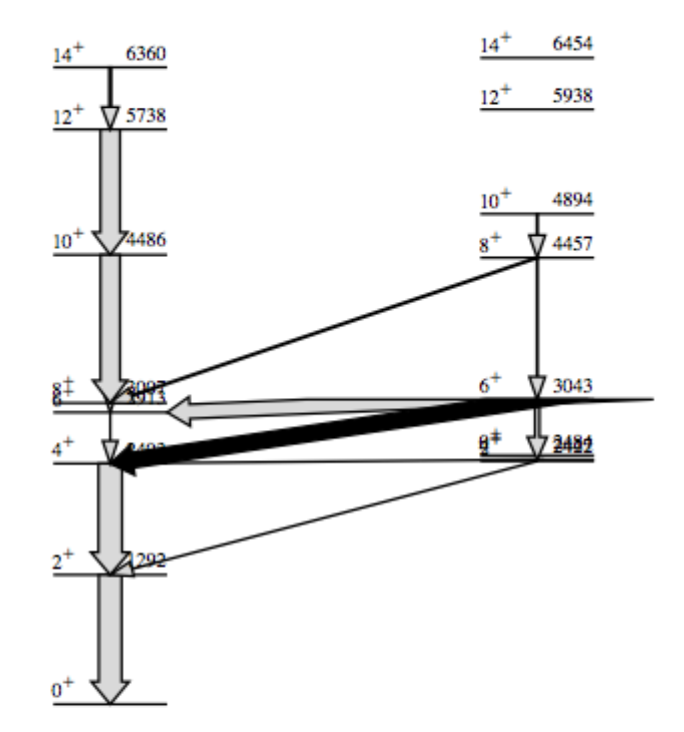}
    \caption{Yrast and Yrare levels $J = 0 - 14$ for $^{92}$Pd using the T0JUN45 effective interaction. B(E2) values are indicated with width corresponding to strength. The Black B(E2:61 $\rightarrow$ 41) is discussed in the text and represents the only strong B(E2) between the displayed levels not shown in Figure \ref{fig1}.}
    \label{fig2}
\end{figure}

    Perhaps yes, perhaps no. By looking beyond the yrast levels in the $T=0$ calculation, we see something that muddies the picture. In Figure 1 we show the calculated levels using JUN45 for the yrast and yrare $J = 0 - 14$ levels and denote any B(E2) over 100 e$^2$fm$^4$ with the width of the arrow indicating the magnitude of the B(E2). In Figure \ref{fig2} we show the calculated levels with T0JUN45. The grey arrows indicate the same B(E2)'s that were shown in Figure \ref{fig1}. The single black arrow represents the one B(E2) over 100 e$^2$fm$^4$ found with T0JUN45 that was absent in the JUN45 calculation for the levels shown.

    That single black arrow shows a strong B(E2)'s from the second $6^{+}$ to the first $4^{+}$ state. This second $6^{+}$ is about 0.1 MeV higher in energy than the lowest $6^{+}$ in the T0JUN45 calculation. It may be that the two lowest $6^{+}$ states have switched order in removing the $T = 0$ matrix elements, allowing us to recapture a strong B(E2:$6 \rightarrow 4)$ without needing to resort to $T = 0$ effects.

    However, one thing clearly missing in the T0JUN45 calculation is a strong B(E2) connecting a $J=8$ to a $J=6$ state. (There is no such transition even if we expand to looking at the first half-dozen $6^{+}$ and $8^{+}$ levels in the T0JUN45 calculation.)

    While the results from \cite{1} would seem to be pointing to the possibility of $T = 0$ pairing, it would appear that the experimental properties of the $J=8$ state are needed to fully settle the question.

    Note that as one goes to higher angular momentum, beyond $J=4$, the B(E2) values steadily decrease. This is typical of what happens in all shell model calculations and indeed experiment. In ref \cite{3} we show the well known formulas for B(E2) values in the rotational and vibrational models. Both show ready increases with $J$. There have been attempts to get rotational bands which terminate e.g the work of Brink et al. \cite{31}. Their work is also described in Igal Talmi's book. \cite{32}.
    
    Qi et al. \cite{33} also point out the importance of looking at electromagnetic transitions. For example they note in the vibrational model the B(E2)'s to the first excited state from  the next $J=0,2,4$  two-phonon triplet should all be the same - $0(2)$ to $2(1)$, $2(2)$ to $2(1)$ and $4(1)$ to $2(1)$. However in their shell model calculations even though the energy levels display a harmonic behavior, the B(E2) $0(2)$ to $2(1)$ is much weaker than the other two.

\section{A Brief Comparison with \texorpdfstring{$^{48}$Cr}{48Cr}}

    In the calculations we have performed \cite{3} both $^{92}$Pd and $^{48}$Cr are treated as eight hole systems with equal numbers of neutrons and protons. For $^{48}$Cr we are at mid-shell in the single $j$ model so that we can also regard this nucleus as consisting of eight particles. Although in the single $J$ space the static quadrupole moment of the $2^{+}$ comes out to be zero, in a full $fp$ space calculation we find a value of $-35.4$ e fm$^2$ -- quite prolate. This is in contrast to the non-mid-shell nucleus Pd where Q(27) is calculated to be $-3.5$.

    It has been noted \cite{10} that at mid-shell we have as a good quantum number the quantity: $sig = (-1)^s$ where $s = \frac{v_{p} + v_{n}}{2}$ and $v_{p}$ and $v_{n}$ are the seniorities of the protons and the neutrons, respectively. See also discussions of the signature selection rule for nuclei with equal numbers of protons as neutron holes by McCullen et al.\cite{8}, Robinson et al. \cite{33} and Lawson \cite{34}. The discussion in Lawson's book is especially illuminating.

    Note that for the yrast even states i.e. $J = 0, 2, 4, ..., 16$ we get an alteration of $sig$. That is to say, $sig > 0$ for $J =4n$ and $sig < 0$ for $J = 4n+2$. Evidently, to get a strong B(E2), a transition between two states of the opposite $sig$ is necessary. For that case the E2 matrix element goes as $e_{p} + e_{n}$ whilst between states of the same $sig$ go as ($e_{p} - e_{n}$). Note that although $sig$ is a good quantum number, in general the total seniority is not.

    An extensive discussion of the symplectic group structure of $N=Z$ nuclei including $^{48}$Cr, $^{88}$Ru, and $^{92}$Pd ground states is given by K. Neergaard \cite{13}. and this group structure is discussed in the work of Talmi \cite{32}.

\begin{figure}[H]
    \centering
    \includegraphics[width=16.5cm]{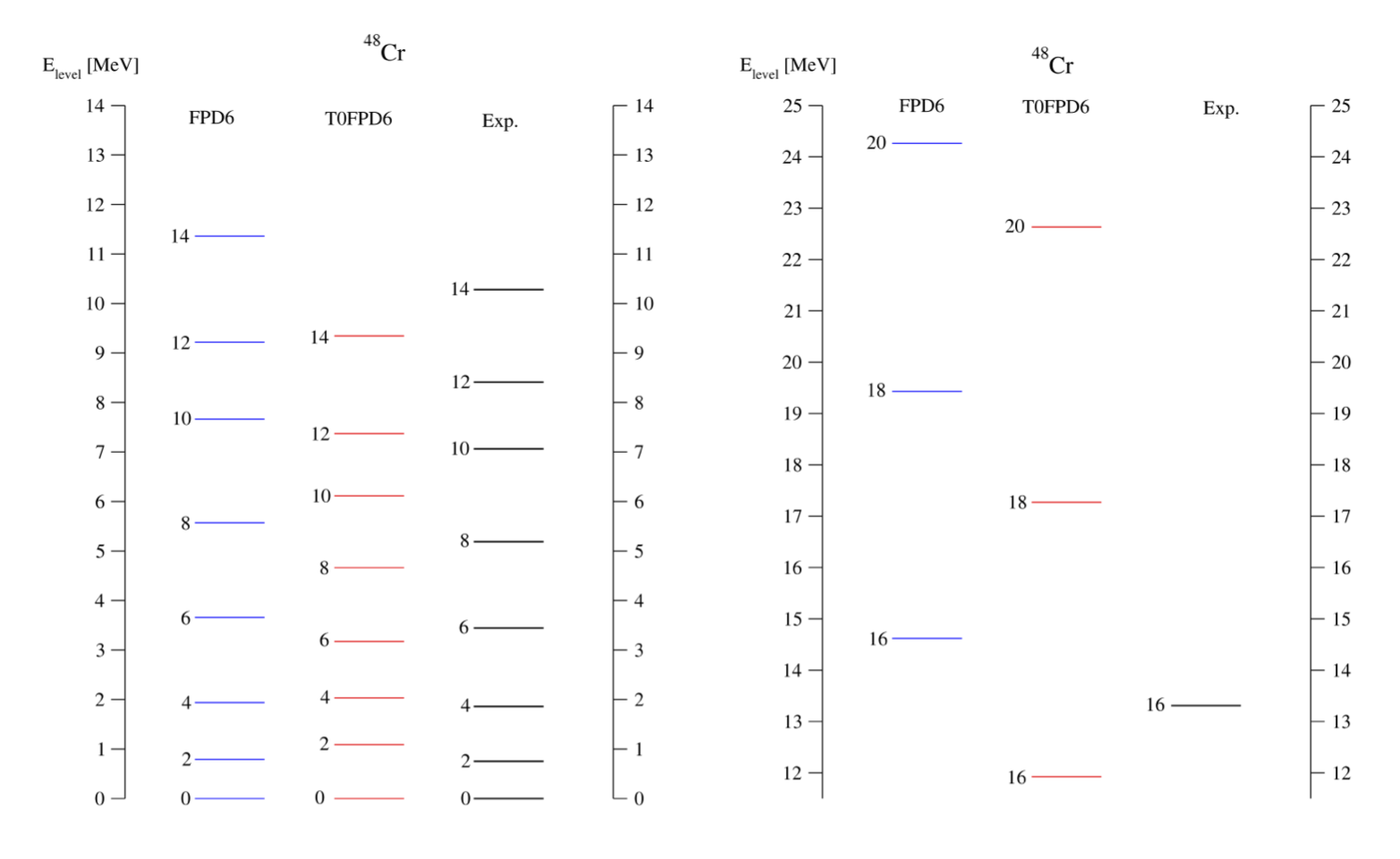}
    \caption{Full $fp$ calculations of even $J$, $T = 0$ states in $^{48}$Cr and comparison with experiment.}
    \label{fig3}
\end{figure}

    For the results of $^{48}$Cr we look at Fig 9 of \cite{23}. Referring to this figure we immediately see a different behavior when we turn the $T = 0$ interaction off in comparison with $^{92}$Pd. In $^{48}$Cr we still get. a more or less equally spaced spectrum, see Figure \ref{fig3}. In fact, it is more equally spaced than when we have the full interaction. One might say that when we turn off the $T = 0$ interaction we go from a weakly rotational spectrum to a more vibrational one. Also there are no catastrophic changes when we turn off the $T = 0$ interaction.

    We have also previously looked at the B(E2) transitions in \cite{26} using the FPD6 interaction with an without the $T = 0$ part of the interaction. As seen in Table \ref{tab2}, for the most part the B(E2)'s with TOFPD6 are smaller than with the full FPD6 interaction. If the ratios were all the same then this could be absorbed by changing the effective charge. Although not quite constant there are no overwhelming red flags as compared with the case of $^{92}$Pd where some of the corresponding ratios are close to zero.

    Hence, in comparing the results for the two nuclei we have shown that when we set $T = 0$ matrix elements to zero we somewhat destroy the linearity in the spectrum of $^{92}$Pd (around $J=6, 8$), but not that of $^{48}$Cr.

    So what is happening? Our calculations indicate that an intruder $6^{+}$ sank below the in band $6^{+}$ state in $^{92}$Pd. We note the second $6^{+}$ state has a strong B(E2) to the first $4^{+}$ state. The B(E2) value is fairly large 122 e$^2$fm$^4$. But there is also a strong B(E2) from the third $6^{+}$ to the first $4^{+}$, 131 e$^2$fm$^4$ in the T0JUN45 results. As mentioned before, the simple B(E2) rules for harmonic vibrator are not in accord with shell model calculations even when the spectra are evenly spaced? Apparently, in $^{48}$Cr there is no such low lying intruder when the $T = 0$ interaction is turned off.

\section{Additional Comments}

    We briefly compare the stretch scheme of Danos and Gilet \cite{35} with Jaz interaction calculations of Escuderos and Zamick \cite{36}. Although at first glance they seem similar there are significant differences. In the former case, it is the wave functions that are stretched whereas in the latter one can say we use a stretched interaction. Since one diagonalizes the Hamiltonian matrix in the latter case, one is assured orthonormal wave functions which satisfy the Pauli principle. In their calculation of $^{44}$Ti wave functions (two valence protons and two valence neutrons) the authors find that they cannot get a $12^{+}$ wave function (the maximum $J$) because the Pauli principle is badly violated. But then an important point is missed-that this state is the ground state when the $J_{max}$ interaction is used. This was actually shown in the context of $^{96}$Cd ($g_{9/2}$ shell) with $I = 16^{+}$ being the ground state \cite{36} but the analogy holds in the $f_{7/2}$ shell.

    There has recently been a reaction against the single $J$ shell picture of the vibrational spectrum in 9 Pd as discussed in ref \cite{1}. Fu, Zhao and Arima \cite{12} state that for $I = 0$ and 2 in $^{92}$Pd the occupancy of $g_{9/2}$ is less than $33\%$. They also quote our ref \cite{3} in this regards. This implies that one will somehow one has to explain the vibrational spectrum in terms of a large space calculation.

\section{Closing Remark}

    We here note that although it easy to see theoretically that the inclusion of odd $J$ pairing changes things relative to its omission, it is easy by examining the limited experimental data to say that one has clear evidence of odd $J$ pairing. Our analysis indicates the $8^{+}$ state is the first one that will clearly distinguish the importance of the components of the $T=0$ interaction, including odd $J$ pairing. We thank Kai Neergaard for useful comments. A more recent paper by Kingan et al. \cite{37} and its relation to ref \cite{27} is also relevant to the considerations that are in this work.

\clearpage
{\renewcommand{\arraystretch}{1.25}
\begin{table}[H]
    \centering
    \caption{Shell Model results for $^{92}$Pd using the JUN45 interaction with and without $T = 0$ two body matrix elements.}
    \begin{tabular}{|c||c|c|c|c|}
        \hline
        $^{92}$Pd & JUN45 & T0JUN45 & Experiment\\
        \hline\hline
        E(2) & 0.840 & 1.292 & 0.874\\
        \hline
        E(4) & 1.721 & 2.406 & 1.786\\
        \hline
        E(6) & 2.515 & 2.913 & 2.535\\
        \hline
        E(8) & 3.217 & 3.007 &  \\
        \hline
        E(10) & 4.070 & 4.486 & \\
        \hline
        E(12) & 4.814 & 5.738 & \\
        \hline
        E(14) & 5.336 & 6.360 & \\
        \hline
        Q(2) & 3.514 & -3.861 &  \\
        \hline
        Q(4) & -7.950 & 1.669 &  \\
        \hline
        Q(6) & -1.868 & 9.307 &  \\
        \hline
        Q(8) & 8.312 & 14.991 &  \\
        \hline
        Q(10) & 7.94 & 18.233 &  \\
        \hline
        Q(12) & 16.406 & 19.139 &  \\
        \hline
        Q(14) & 25.892 & 29.929 &  \\
        \hline
        B(E2:2$\rightarrow$0) & 304.5 & 182.3 &  \\
        \hline
        B(E2:4$\rightarrow$2) & 382.6 & 236.7 & \\
        \hline
        B(E2:6$\rightarrow$4) & 364.1 & 5.018 &  \\
        \hline
        B(E2:8$\rightarrow$6) & 315.1 & 2.474 &  \\
        \hline
        B(E2:10$\rightarrow$8) & 334.8 & 155.7 &  \\
        \hline
        B(E2:12$\rightarrow$10) & 304.6 & 158.2 &  \\
        \hline
        B(E2:14$\rightarrow$12) & 250.0 & 15.607 &  \\
        \hline
    \end{tabular}
    \label{tab1}
\end{table}}

{\renewcommand{\arraystretch}{1.25}
\begin{table}[H]
    \centering
    \caption{Ratio of $^{48}$Cr yrast B(E2) values [e$^2$fm$^4$] in full FPD6 and T0FPD6.}
    \begin{tabular}{|p{2cm}|p{6cm}|p{6cm}|}
    \hline
    Transition & Ratio of B(E2) transitions with and without the $T=0$ effective interaction & Corresponding ratio in $^{92}$Pd\\
    \hline\hline
    2$\rightarrow$0 & 0.590 & 0.598\\
    4$\rightarrow$2 & 0.543 & 0.618\\
    6$\rightarrow$4 & 0.399 & 0.013\\
    8$\rightarrow$6 & 0.491 & 0.008\\
    10$\rightarrow$8 & 0.523 & 0.465\\
    12$\rightarrow$10 & 0.784 & 0.519\\
    14$\rightarrow$12 & 0.803 & 0.06\\
    16$\rightarrow$14 & 1.029 & \\
    18$\rightarrow$16 & 0.815 & \\
    20$\rightarrow$18 & 0.247 & \\
    \hline
    \end{tabular}
    \label{tab2}
\end{table}}

\end{document}